\documentclass[aps,prl,twocolumn,groupedaddress,showpacs,showkeys]{revtex4-1}

\usepackage{verbatim}
\usepackage{amsmath}
\usepackage{graphicx}
\usepackage{array}
\usepackage{SIunits}
\usepackage{color}
\usepackage{float}
\usepackage{placeins}
\usepackage{epstopdf}
\usepackage{upgreek}
\usepackage{natbib}

\begin{document}


\title{Broadband and efficient Quantum Memory Using ac Stark Gradient Echo Memory}


\author{Mahmood Sabooni$^{a,b}$}
\email[]{msabooni@uwaterloo.ca}
\author{Mohsen Jafarbeklu$^{b}$}
\author{Farrokh Sarreshtehdari$^{b}$}

\affiliation{$^{a}$ Institute for Quantum Computing, Department of Physics and Astronomy, University of Waterloo, Waterloo, Ontario, N2L 3G1, Canada.}
\affiliation{$^{b}$ Department of Physics, University of Tehran, 14399-55961, Tehran, Iran.}


\date{\today}

\begin{abstract}
A quantum state light-storage, using a virtual magnetic field through the ac Stark effect is proposed to combine the high overall storage efficiency and large bandwidth employing room temperature atomic vapor. In this approach, which was called the ac Stark Gradient Echo Memory (ASGEM), it has been shown the possibility to employ about a nanosecond ac Stark pulse far detuned ($\sim 127$ THz) from $D_1$ line of rubidium and create an atomic media with the possibility to store a photon with about a GHz bandwidth with storage and retrieval efficiency of more than $ 90\%$. A contour plot of efficiency as a function of gradient field strength and optical depth, based on three-level Maxwell-Bloch equations, simulated for a better understanding of experimental parameter optimization.

\end{abstract}


\maketitle


\par
In the quantum world, the phase angles between the components of a system in a quantum superposition needs to be well defined, otherwise the quantum state will decohere. Keeping track of phase information is required to maintain ``quantumness", and is what makes quantum systems different from systems in the classical world. In the case of a single photon, the photon plays the role of a flying qubit and needs to maintain ``quantumness". In addition, the requirement of having a stationary qubit (like matter) and map on the encoded quantum state from flying qubit to stationary qubit is perceived. Such a quantum interface between light and matter, with an ability to map onto, store in, and later retrieve the quantum state of light from matter, is vital in quantum information processing \cite{Hammerer2010} and is called a quantum memory for light, which is the main target of this paper.

\par
There are several proposals which can be realized based on presence of a functional quantum memory for light in the quantum information community. Quantum computing based on linear optics \cite{Kok2007}, signal synchronization in optical quantum computation \cite{Knill2001}, implementation of a deterministic single-photon source \cite{Chen2006,Eisaman2011}, quantum transduction between an optical to microwave photon for hybrid quantum computation \cite{Kurizki2015}, and long-distance quantum communication through the concept of quantum repeaters are some examples \cite{Bennett1993,Duan2001,Waks2002,Boaron2018,Lucamarini2018}. Bring all mentioned applications to reality are decent motivations towards research investment for realization of full functional quantum memories for light.
\par
\begin{figure}[t]
    \includegraphics[width=8cm]{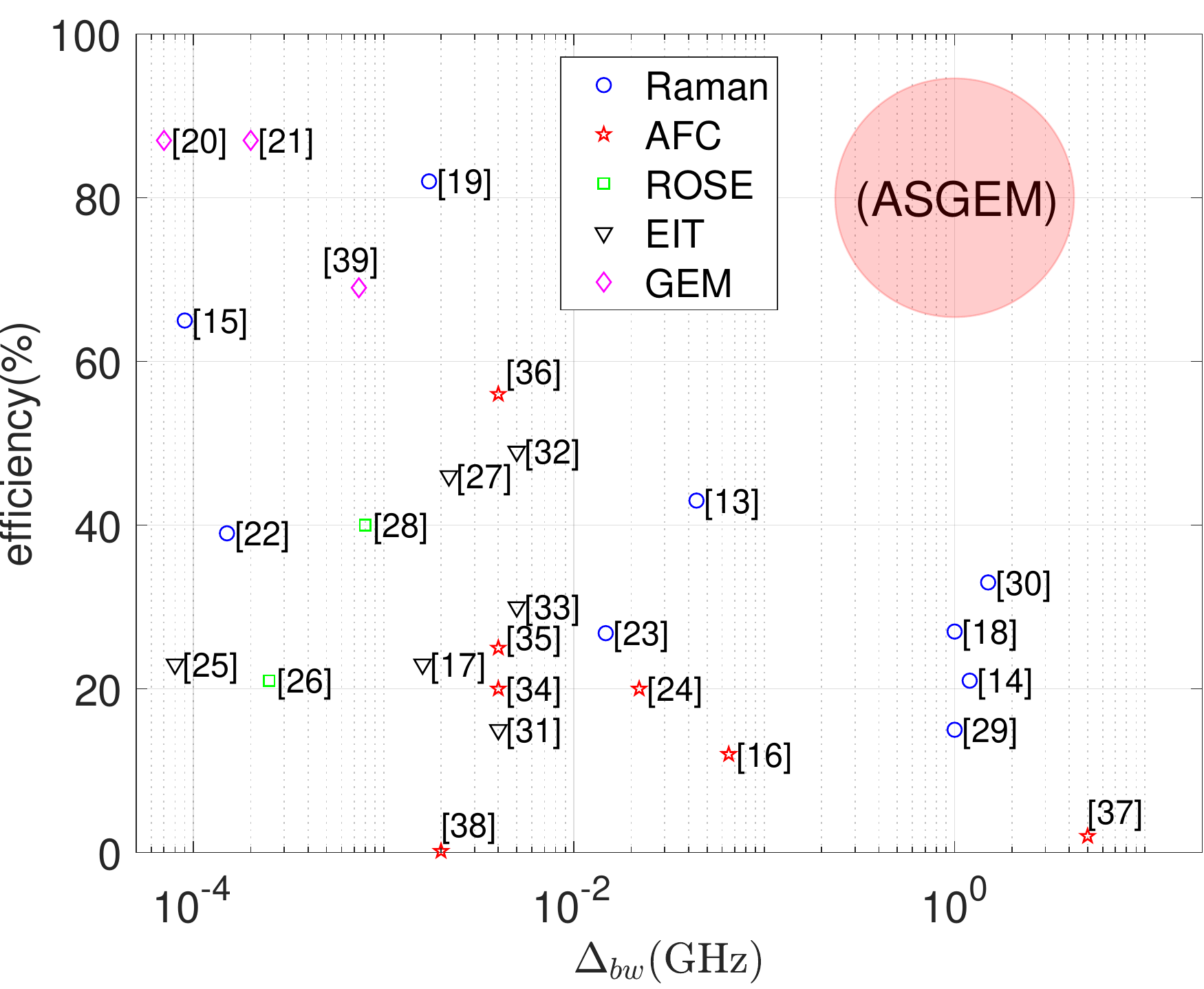}
    \caption{ (Color online) The state-of-art for the bandwidth and overall storage and retrieval efficiency for ensemble based quantum memory for light \cite{Thomas2019,Michelberger2015,Vernaz-Gris2018,Akhmedzhanov2016,Rastogi2019,Sprague2014,Guo2019,Cho2016,Hosseini2011,Kutluer2016,Ding2016,Zhou2015,Blatt2016,Minnegaliev2018,Hsiao2018,Dajczgewand2014,Reim2010,Reim2011,Nicolas2014,Zhou2012,Riedl2012,Maring2014,sabooni2010,sabooni2013,Saglamyurek2011,Park2018,Hedges2010}. The aim of this paper is to combine these two properties in a single physical setup as illustrated the shaded area called ASGEM . Theoretical limitation*: The overall efficiency of all photon echo based quantum memory protocols running in the forward direction suffer from a theoretical limit which called re-absorption \cite{Sangouard2007}. This limitation was solved by employing a gradient magnetic field along the sample in gradient echo memory (GEM) and by placing the ensemble inside a low-finesse cavity in atomic frequency comb (AFC). }
    \label{fig1}
\end{figure}

Based on the target application, different properties of a quantum memory need to be optimized. The fidelity which is the quantity characterizes the similarity between the stored and retrieved quantum state \cite{Nielsen2000}. The overall storage and retrieval efficiency is simply defined as the energy of the pulse recalled from the memory divided by the energy of the pulse sent in for the storage. The storage time, bandwidth of the memory and operational wavelength are other important properties. The state-of-art for the bandwidth and overall storage and retrieval efficiency for ensemble based quantum memory for light is illustrated in Fig. \ref{fig1}. The aim of this paper is to propose an approach, called ac Stark gradient echo memory (ASGEM), and improve the quantum memory bandwidth and the overall efficiency in a single physical setup as illustrated by the shaded circle in Fig. \ref{fig1}. 
\par
The warm atomic vapor cell is employed to achieve the world record storage and retrieval efficiency (87\% \cite{Hosseini2011}) for a few MHz photon employing gradient echo memory (GEM) technique as shown in Fig. \ref{fig1}. On the other hand, a GHz photon is stored, in the warm atomic vapor cell, with an efficiency of $< 30\%$ \cite{Reim2010,Michelberger2015} employing off-resonance Raman technique which is almost approaching its limit because of the re-absorption of the signal in the rest of the sample in the forward direction. It should be mentioned that in the calculation of storage and retrieval efficiency in Refs. \cite{Reim2010,Michelberger2015} the filtering stages after the memory, which is essential to discriminate output single photon from bright control pulse, are excluded. Extending the bandwidth of the memory in a conventional GEM technique from MHz to GHz requires applying a large magnetic field and possibility to switch within several nano-second which is not possible with current technologies. We propose employing virtual magnetic field instead of conventional magnetic field. This is possible through applying a far detuned laser field which can create an ac Stark shift for the target transition. We have the possibility to switch the virtual magnetic field much faster than the conventional magnetic field. Based on this proposal, one can build a full broadband
electromagnetic-controlled quantum memory using an ac Stark shift instead of the conventional magnetic field in the GEM technique. Improving quantum memory bandwidth has some specific benefit for instance for single photon storage emitted from quantum dots.

\par
The dephasing-rephasing process in photon echo based techniques, like GEM, is done via applying electric field (in solids) or magnetic field (in vapors) with possibility to switch the direction of the field at some particular point of the storage sequence. The gradient field proposed in GEM makes it unique since it is possible to make all atoms in the rest of the sample off-resonance with the emitted echo from earlier slices in the sample. This gradient field along the propagation direction will suppress the re-absorption problem. The slope of the gradient field and the sample length will define the bandwidth of the memory. Approaching GHz band-width quantum memory employing conventional magnetic field (conventional GEM approach) is limited to the possibility of fast switching (nano-second) of large magnetic fields.

\par
\begin{figure*}[t!]
$\begin{array}{rl}
    \includegraphics[width=0.4\textwidth]{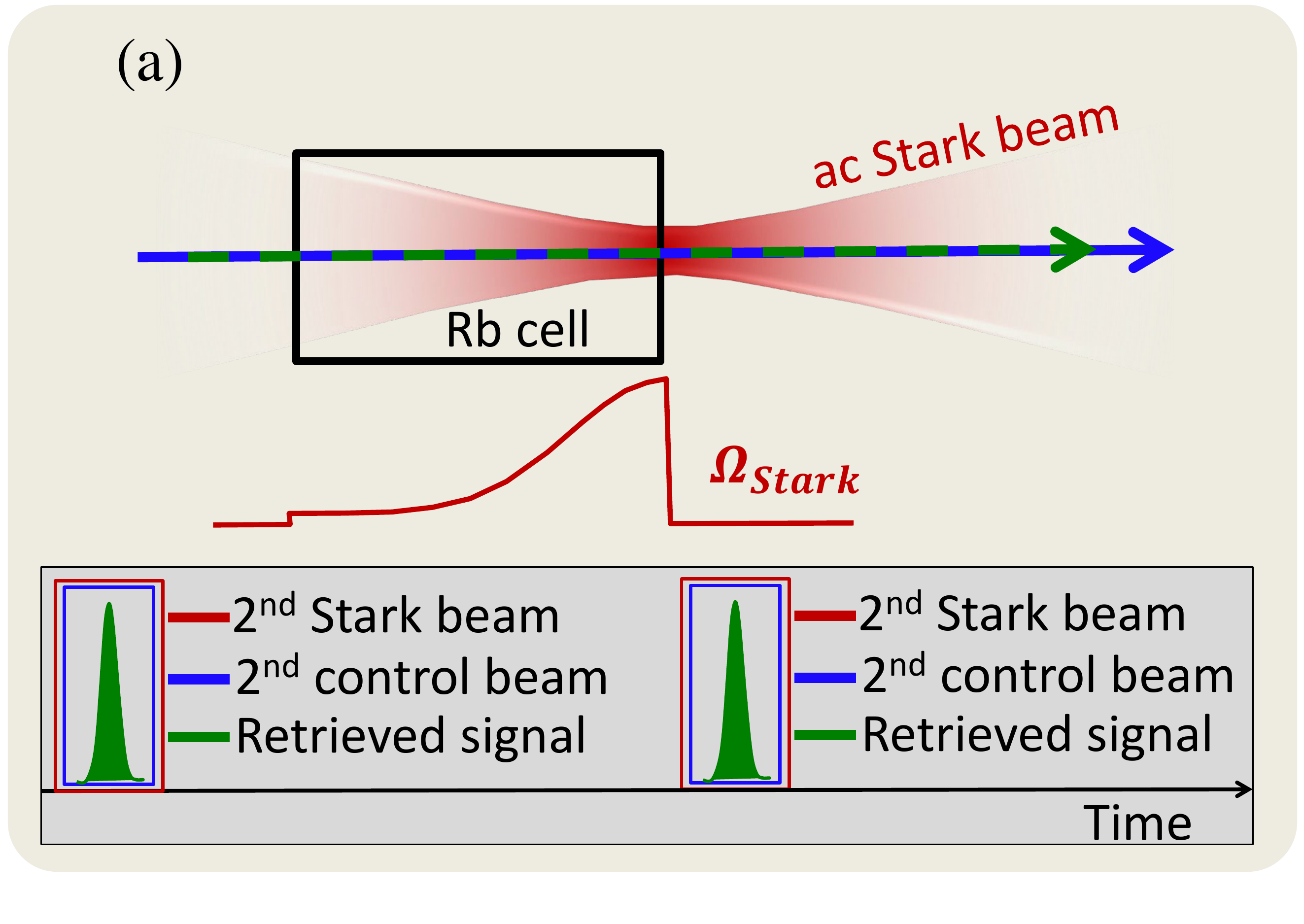} &
    \includegraphics[width=0.5\textwidth]{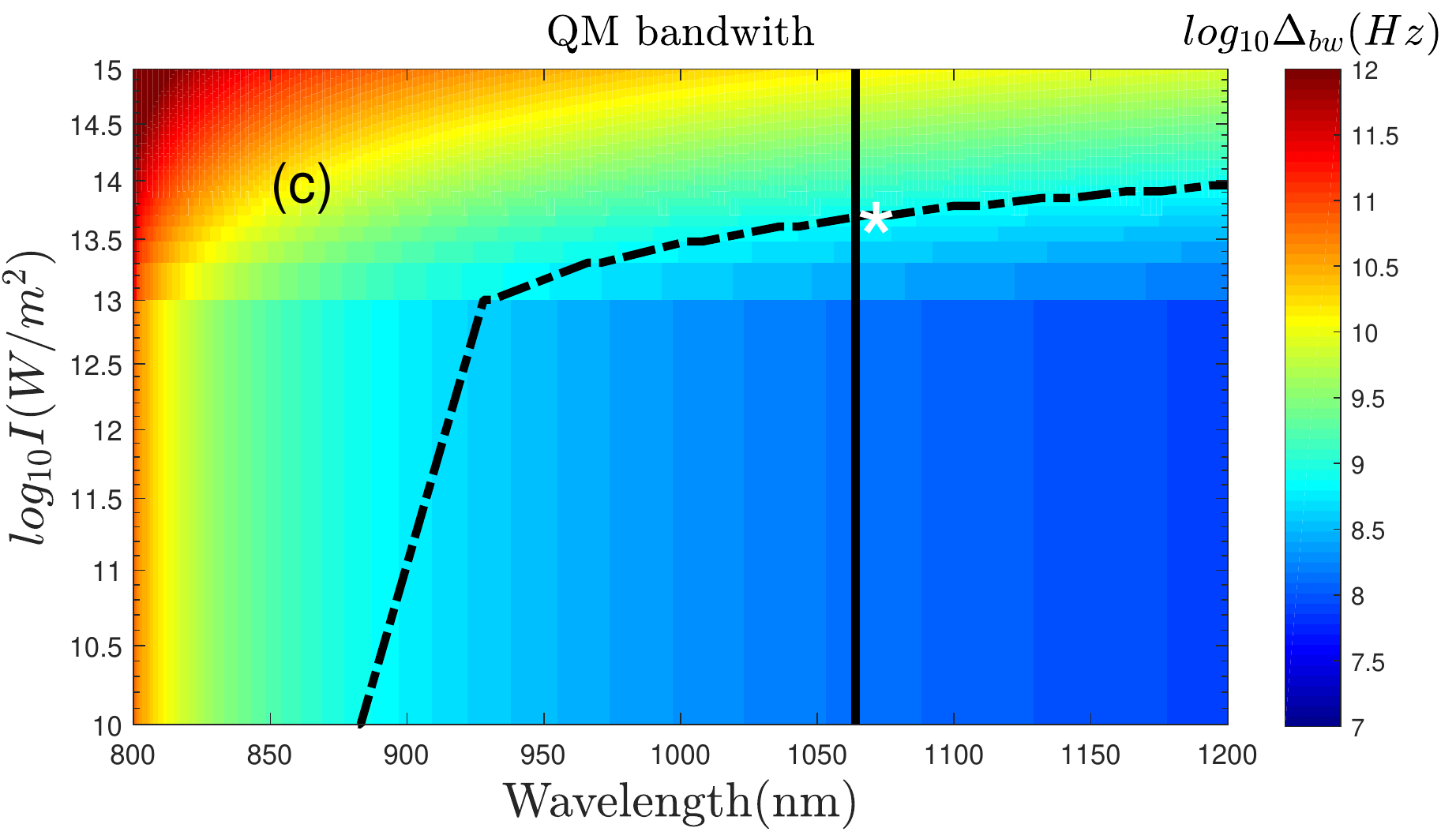} \\

    \includegraphics[width=0.4\textwidth]{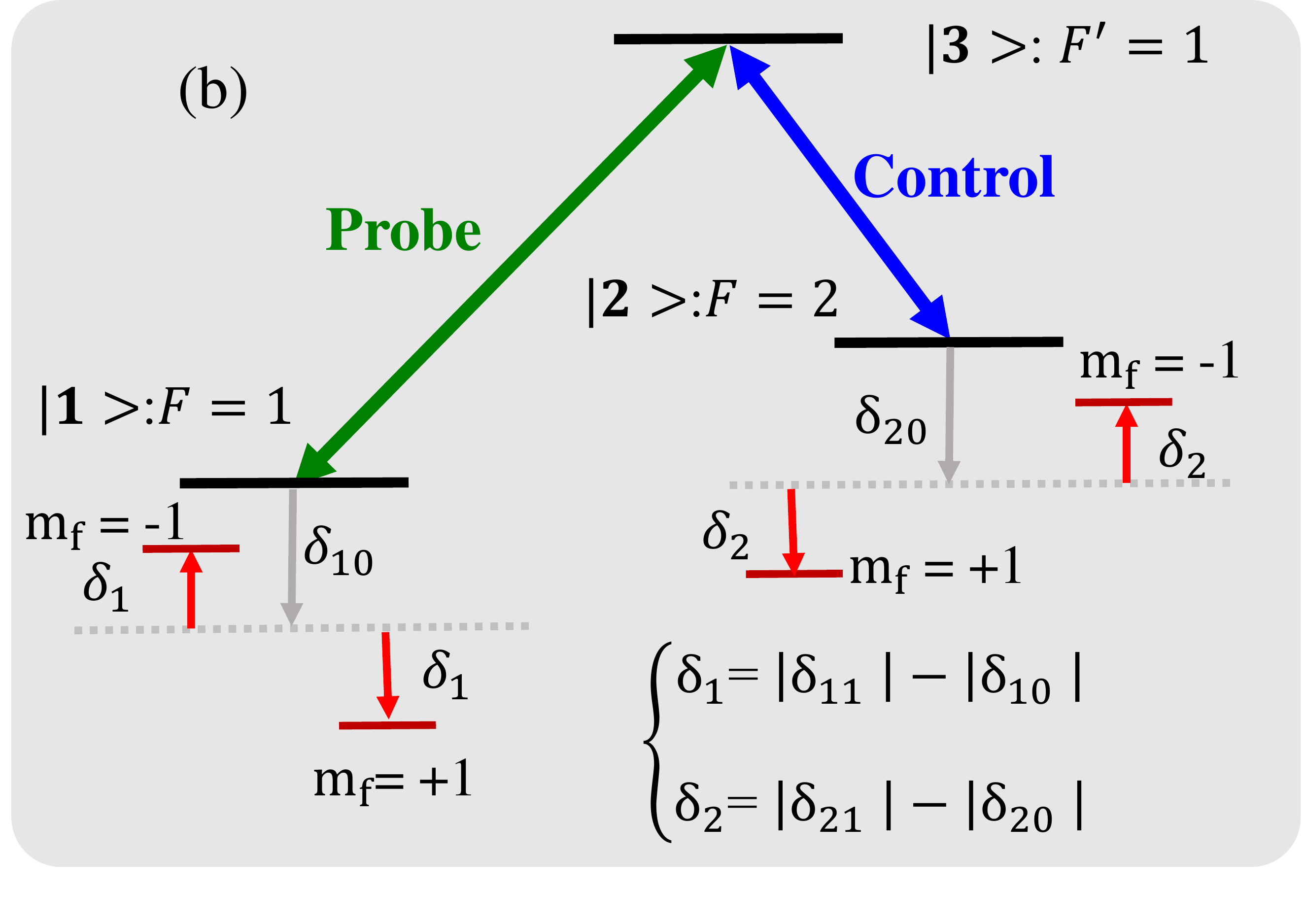}&
    \includegraphics[width=0.5\textwidth]{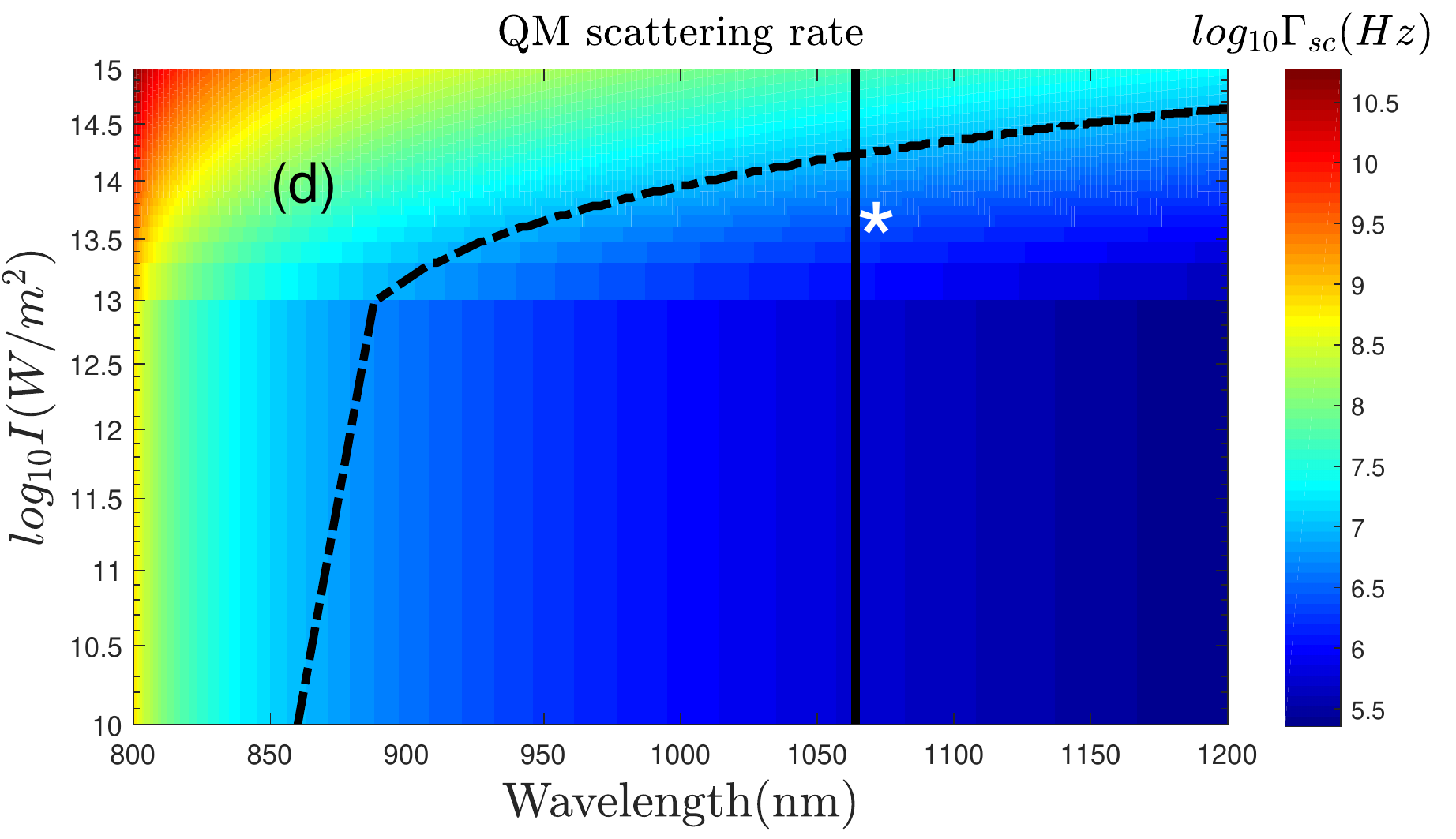} \\
\end{array}$
\caption {(Color online) (a) A tight focused ac Stark beam is employed to prepare a gradient virtual magnetic field along the sample. A bright control field and single photon level signal field in a $\Lambda$-scheme configuration employed for quantum state storage. Experimental sequence. Switching the polarization or applying a phase shift of $\pi$ between first and second ac Stark beam can reverse the gradient along the sample. (b) Schematic for total energy shift of ground states produced via ac Stark field. $^{87}Rb$ ground states energy shift produced via ac Stark field (c) Contur plot of  quantum memory bandwidth $\Delta_{bw}$ for different wavelength and ac Stark beam intensity. The dashed line is the border for $\Delta_{bw}=1$GHz. (d) Contour plot of scattering rate $\Gamma_{sc}$ for different wavelength and ac Stark beam intensity. The dashed line is the border for $\Gamma_{sc}=5$MHz. In both c and d the solid line represent the $\lambda=1064$ nm. The asterik is representing the ac Strak intensity of $5\times10^{13}\frac{W}{m^2}$ producing $\Delta_{bw}=1$GHz and $\Gamma_{sc}=1$MHz.  \label{fig:Main}}
\end{figure*}

The main purpose of this proposal is to employ a virtual magnetic field via sending a detuned (several tens of THz) light pulse into the sample. Considering the laser intensity gradient along the propagation axis, one can create a different ac Stark shift of the atomic transitions along the sample. Borrowing the idea proposed by Sparkes et al. \cite{Sparkes2010} and employing a pulsed laser, the memory bandwidth will increase and become suitable for storage of photon emitted from semiconductor quantum dot. The ac Stark beam needs to be on during the writing into and reading out of the memory for about several nanoseconds which is shown in Fig. \ref{fig:Main}a.\\

\par
The idea behind optical dipole potential \cite{Grimm2000} is borrowed to calculate ac Stark energy level shift $\Delta E_g$ and the scattering rate $\Gamma_{sc}$ produced by ac Stark beam. The effect of far-detuned light on atomic levels could be treated based on second order time-independent perturbation theory as follows: $\Delta E = \sum\limits_{u\neq p} \frac{|\langle p|H_1|u\rangle|^2}{\varepsilon_u-\varepsilon_p}$
where $H_1=-\hat{\mu}.\textbf{E}$ with $\hat{\mu}=-e\textbf{r}$ is the electric dipole moment, which take into account the coupling strength between atomic sub-levels and $\textbf{E}=E_0\hat{\textbf{e}}_q$ which is the electric field of ac Stark beam with polarization direction q. In the simple form, the unperturbed state could be considered as $n$ photon with total energy $\varepsilon_u=n\hbar\omega_l$. When one photon is absorbed by the atom at angular frequency of $\omega_r$ the total energy of atom plus field (perturbed state) will be $\varepsilon_p=\hbar\omega_r+(n-1)\hbar\omega_l=-\hbar\Delta_{pu}+n\hbar\omega_l$. Now, one can conclude: $\varepsilon_u-\varepsilon_p=\hbar\Delta_{pu}=\hbar(\omega_l-\omega_r)$.
Considering $I=2\epsilon_0c|\textbf{E}|^2$, the energy shift of the ground state will be \cite{Grimm2000}:
\begin{equation} {\label{eq:shift}}
\Delta E_g = \frac{I}{2\hbar\epsilon_0c} \sum\limits_{i} \frac{|\langle i|\hat{\mu}.\hat{e}_q|g\rangle|^2}{\omega_l-\omega_{gi}}
\end{equation}
For states $|g\rangle=|J,F,m_F\rangle$ and $|i\rangle=|J',F',m_{F'}\rangle$, the matrix element can be broken down as follows:
\begin{widetext}
\begin{equation} {\label{eq:CG1}}
\langle F,m_F|er_q|F',m_{F'}\rangle=\langle F||er_q||F'\rangle(-1)^{F'-1+m_F}\sqrt{2F+1}
\begin{pmatrix}
 F' & 1 & F  \\
 m_F' & q & -m_F  \\
\end{pmatrix}
\end{equation}
\begin{equation} {\label{eq:CG2}}
=\langle J||er_q||J'\rangle(-1)^{F'+J+I+1}\sqrt{(2F+1)(2F'+1)(2J+1)}
\begin{Bmatrix}
 J & J' & 1  \\
 F' & F & I  \\
\end{Bmatrix}
\begin{pmatrix}
 F' & 1 & F  \\
 m_F' & q & -m_F  \\
\end{pmatrix}
\end{equation}
\end{widetext}
where the ac Stark field assumed to be fully polarized along one spherical component $q$ ( 0 stands for linear polarization while $\pm1$ is representing right or left circular). The final equation for the light shift imposed onto a given ground state $|J,F,m_F\rangle$ by Stark beam with intensity $I$ detuned by $\Delta_{FF'}$ from all possible states of the atoms, could be derived as follows:
\begin{widetext}
\begin{equation} {\label{eq:Final_Shift2}}
\Delta E_g=\frac{I}{2\hbar\epsilon_0 c}\sum\limits_{F\neq F',m_F'\neq m_F}\frac{|\langle J||er_q||J'\rangle|^2}{\omega_l-\omega_{FF'}} (2F+1)(2F'+1)(2J+1)
\begin{Bmatrix}
 J & J' & 1  \\
 F' & F & I  \\
\end{Bmatrix}^2
\begin{pmatrix}
 F' & 1 & F  \\
 m_F' & q & -m_F  \\
\end{pmatrix}^2
\end{equation}
\end{widetext}
\par
In addition to the energy shift of the atomic structure produced via ac Stark beam, there is a possibility for the incoming photon to be absorbed and re-emitted. This process, which is called scattering, can affects atomic coherence time and finally limit the quantum memory storage time. The scattering rate $\Gamma_{sc}$ for a given ground state $|g\rangle$ is connected to the energy shift calculated in Eq. \ref{eq:shift} as follows: $\hbar\Gamma_{sc}=\frac{\Gamma}{\Delta} (\Delta E_g)$
where $\Gamma$ is the damping rate of the re-emitted photon, and $\Delta$ is the detuning $\omega_l-\omega_{gi}$ \cite{Grimm2000}.
\par
In general, the damping rate $\Gamma$ (corresponding to the spontaneous decay rate of the excited state) can be determined by the dipole moment transition between ground and excited states:
$\Gamma_{sc}=\frac{\omega^3_l}{3\pi\epsilon_0\hbar c^3}|\langle f|e\hat{\mathbf{r}}.\mathbf{\epsilon}_{q_{sc}}|i\rangle|^2$.
The final equation for the scattering rate $\Gamma_{sc}$ produced via ac Stark beam will be:
\begin{equation} {\label{eq:scattering_rate_final}}
\Gamma_{sc}=\frac{I\omega^3_l}{6\pi\epsilon^2_0\hbar^3 c^4}\sum\limits_{i}|\frac{\langle f|e\hat{\mathbf{r}}.\mathbf{\epsilon}_{q_{sc}}|i\rangle\langle i|e\hat{\mathbf{r}}.\epsilon_q|g\rangle}{\omega_l-\omega_{gi}}|^2
\end{equation}
\begin{figure*}[t!]
$\begin{array}{rl}
    \includegraphics[width=0.5\textwidth]{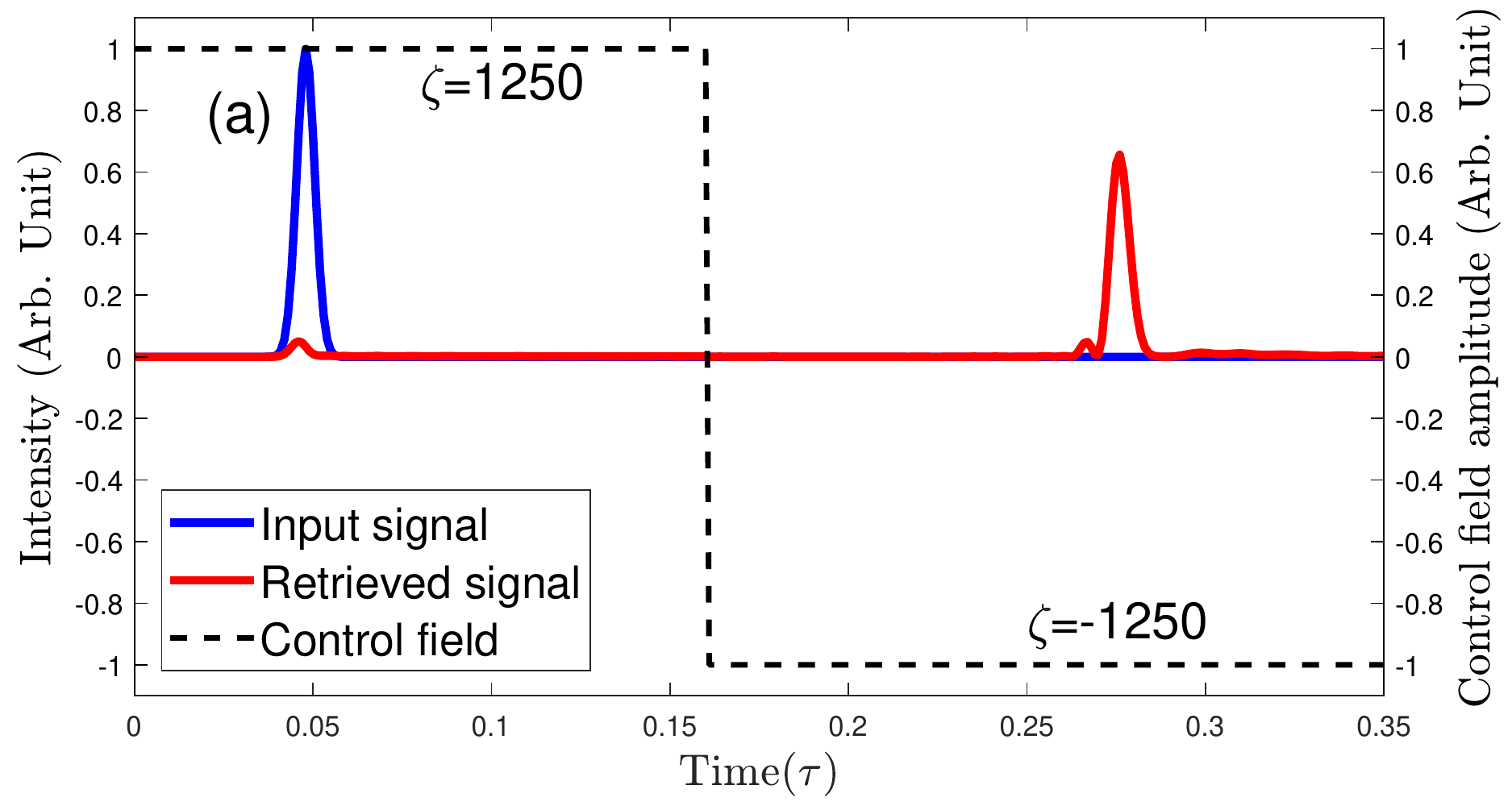} %
    \includegraphics[width=0.5\textwidth]{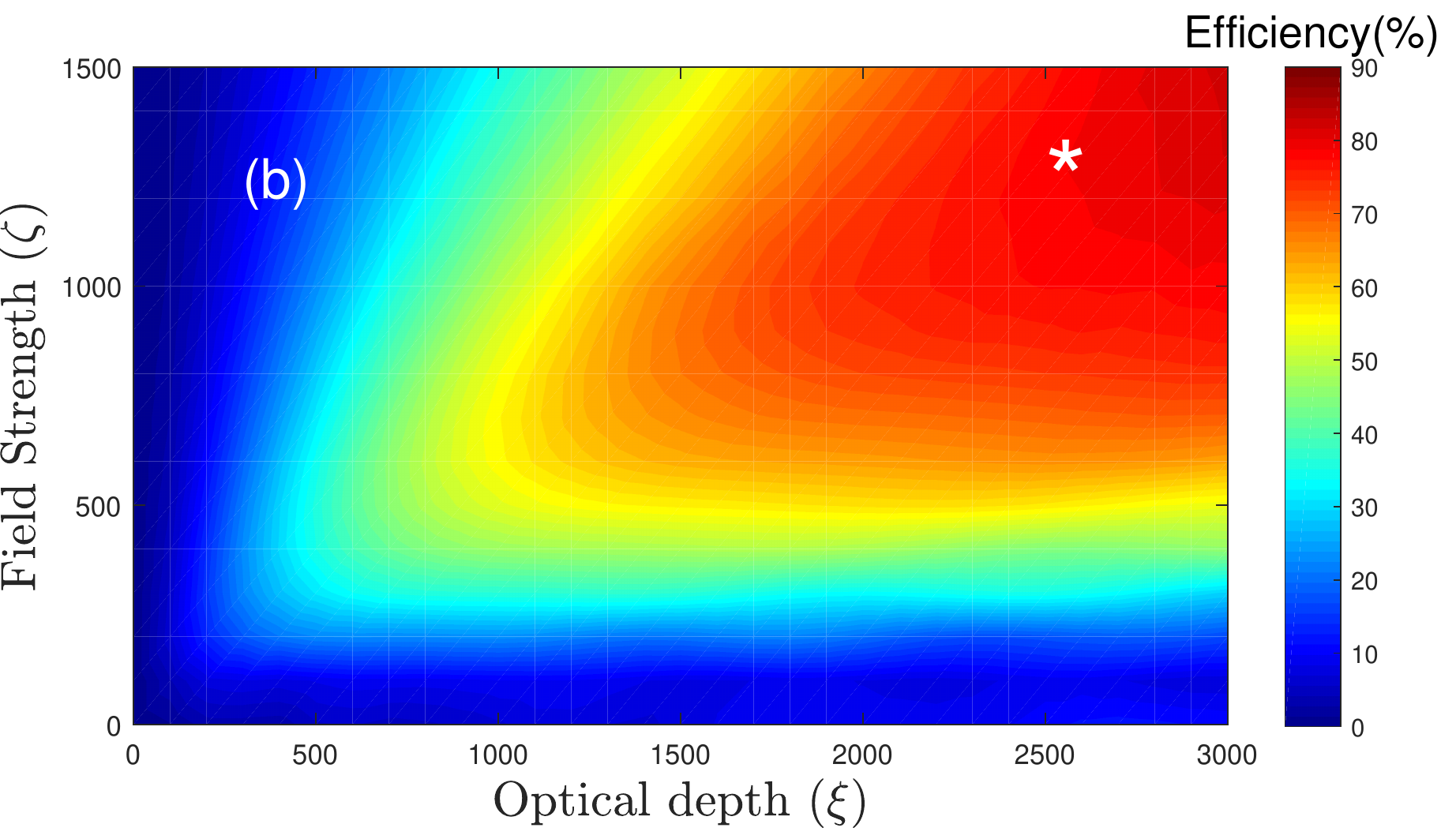} \\
\end{array}$
\caption {(Color online) (a) The control field (black dashed line, right axis) is reversed $(\zeta \rightarrow -\zeta)$ at $t=0.16\tau$, and the stored pulse (blue solid line) will be retrieved as an echo signal  (red solid line) at $t=0.28\tau$ for $(\xi=2500,\zeta=1250)$. The efficiency for this case is around 75$\%$. This is shown in Fig.3bby an asterisk. (b) Contour plot of efficiency $R(\xi,\zeta)$ as a function of gradient filed strength $\zeta$ and optical depth $\xi$. The efficiency for $(\xi=2500,\zeta=1250)$ is shown by a whiasterisk in the contour plot.
	\label{fig9}}
\end{figure*}
\par
The optimal intensity distribution of the ac Stark shift beam could be designed in the way that two beams from two sides of the sample impinging on the cell. One of them could be set to have the maximum intensity in the beginning of the sample while the second one should have the maximum intensity at the end of the sample. As an alternative, the ac Stark beam can propagate along the sample while it sets very un-collimated and the focal point is at the back side of the sample as illustrated in Fig.\ref{fig:Main}a. The direction of the gradient slope can change via changing the polarizations of the beam since the direction of Stark shift of the atomic levels are polarization dependent. The gradient slop can be changed also by applying a phase shift of $\pi$ between the first (write) and second (read) ac Stark fields \cite{Liao2014}.

\par
The $D_1$ line of rubidium atoms ($^{87}Rb$) are one of the promising candidates. Following the Eq. \ref{eq:Final_Shift2}, we employed the transition dipole matrix of $\langle J=1/2||er||J'=1/2\rangle$ which is $2.5377(17).10^{-29} C.m $ and perform summation over all possible $|F,m_F\rangle$ states \cite{Steck2001}. The ac Stark intensity $I$ and detuning $\Delta$ estimated to produce a $\Delta_{bw}\sim$GHz wide bandwidth memory. Following the $\Lambda-$ type protocols, two specific ground states $|1,m_F\rangle$ and $|2,m_F\rangle$ is employed to calculate the relative energy shift of these two ground states to exceed our target bandwidth ($\sim$ GHz) as shown in Fig. \ref{fig:Main}b. Chasing our convention in Fig. \ref{fig:Main}b, one can derive the maximum total broadening and therefore the memory bandwidth to be as follows:
$\Delta_{bw}=|\delta_{20}-\delta_{10}|+|\delta_1|+|\delta_2|$ where $\delta_{F,m_F}$ is representing the energy shift of every single $|F m_F\rangle$ state compared to the original (no ac Stark beam) $|F,m_F=0\rangle$ state. Demonstration of a GHz bandwidth memory, using ac Stark pulsed laser ($\sim 2ns$) at 1064 nm and beam diameter of $10 \mu m$ needs intensity of about $5\times10^{13} W/m^2$ at focus which is energy of about $500\mu J$ (see Figs. \ref{fig:Main}c).\\

\par
The scattered photon events should be lower than the atomic decay rate during the dephasing-rephasing process. Suppose the target transition for quantum state storage is $D_1$ line of $^{87}Rb$ while the Stark beam is detuned by $\Delta$ from the target transition. The scattering rate is scaled as $I/\Delta^2$ while the ac Stark shift is scaled as $I/\Delta$ where $I$ is the ac Stark beam intensity. Therefore, employing Eq. \ref{eq:scattering_rate_final} and increasing the ac Stark beam detuning (up to 127 THz) will help to suppress the scattering rate to the acceptable level ($<\Gamma=5$ MHz) (see Figs. \ref{fig:Main}d).

\par
The storage and retrieval dynamics in our protocol could be simulated employing three level Maxwell-Bloch equations in the region of $|\Omega_p|\ll\Gamma$ as follows \cite{Liao2014}:
\begin{equation} {\label{eq: MB equations}}
\begin{aligned}
&\partial_{t}\rho_{31}= -\left(\frac{\Gamma}{2}+ i\Delta_{p}\right) \rho_{31} + \frac{i}{2} \Omega_{c} \rho_{21} + \frac{i}{2} \Omega_{p}, \\
&\partial_{t}\rho_{21}= i(\Delta_{c}-\Delta_{p}+i\gamma)\rho_{21} + \frac{i}{2} \Omega_{c}^{*} \rho_{31}, \\
&(\partial_{z}+\frac{1}{c} \partial_{t})  \Omega_{p}= i \eta \rho_{31}.
\end{aligned}
\end{equation}

Where in our case we have employed the $D_1$ line $(5^2s_{1/2}\rightarrow5^2p_{1/2})$ of $^{87}Rb$ with $\Gamma \sim 5$ MHz representing the spontaneous decay rate of the excited state $|3\rangle$ and $\gamma$ is the decoherence rate between two ground states. $\Delta_c (\Delta_p) = 50 (200)$ MHz is the detuning of the control (probe) field. In addition, $\Omega_c=500$ MHz and $\Omega_p=5$ kHz  are Rabi frequencies of control and probe fields, respectively. Furthermore, $\eta$ is defined as $\frac{\Gamma \xi}{2L}$ where $\xi$ and $L$ are the optical depth and the length of medium, respectively.
\par
One could consider a case in which the gradient control field changes linearly as  $\Omega_c(z)= \frac{\zeta \Gamma z }{L}$ over length of the medium where $\zeta$ is the strength factor of the gradient field. A broadband Gaussian pulse $\Omega_p(t)= \Omega_{p0} e^{-(t-t_0/\kappa)^2}$ with duration of $1$ns ($\kappa = 0.005\tau$) with a center at $t_0=0.048\tau$ enters the medium.
All time spectra are represented in the unit of the lifetime $\tau$ of the excited state $|3>$.

Then by reversing the control field $(\zeta \rightarrow -\zeta)$ at $t=0.16\tau$, the stored pulse will be retrieved as an echo signal at $t=0.28\tau$ with the width same as initial pulse (see Fig. \ref{fig9}a). The storage efficiency of the quantum memory can be obtained by calculating the ratio of retrieved energy to initial pulse energy as follows:$R(\xi,\zeta)= (\int_{0.16\tau}^{\infty} |\Omega_{p}(t,L)|^2  dt)/(\int_{0}^{\infty}|\Omega_{p}(t,0)|^2  dt)$ where as we mentioned earlier $0.16\tau$ is the time we apply a $\pi$ phase shift to $\Omega_c$ by reversing the gradient control field which leads to the generation of echo signal eventually. As an example, we consider a case in which we have $\xi=2250$ and $\zeta=1250$. The resultant echo with the aforesaid conditions is illustrated in Fig. \ref{fig9}a with calculated efficiency of about $78\%$. In addition, as shown in a contour plot of Fig. \ref{fig9}b, one can improve storage and retrieval efficiency to even higher than $90\%$.





\par
In conclusion, the optimization, manipulation, and control of broadband quantum state are an important task for fundamental and applied physics. Quantum state storage is one of the vital components for the development of many devices in quantum information processing. As discussed earlier the main issue in the quantum memory community is to combine the state-of-the-art of properties in a single physical system. This is an absolutely vital step to bring the quantum memory to reality. We have proposed a technique to combine the high efficiency and large bandwidth properties in a single physical system which can be operated at room temperature. The quantum state storage optimization is done in the way that the platform can be employed to perform a hybrid experiment between atomic system (rubidium) and semiconductor quantum dot. We have shown the possibility to employ $\sim$ nanosecond ac Stark pulse far detuned ($\sim 127$ THz) from $D_1$ line of rubidium and create an atomic media with possibility to store a photon with $\sim$ GHz bandwidth with storage and retrieval efficiency of $\sim 90\%$.

\par
We gratefully acknowledge the Iran National Science Foundation (INSF) for the financial support of this study (Grant No. 96003886).

\end{document}